\documentclass[epjCONF]{svjour}
\usepackage{graphics}
\usepackage[varg]{txfonts} 
\usepackage[latin1]{inputenc}
\session-title{New technologies for probing the diversity of brown dwarfs and exoplanets}
\begin{document}
\title{Consequences of spectrograph illumination for the accuracy of radial-velocimetry}
\author{Boisse, I.\inst{1}\fnmsep\thanks{\email{iboisse@iap.fr}} \and Bouchy, F.\inst{1, 2} \and 
Chazelas, B.\inst{3} \and Perruchot, S.\inst{2} \and Pepe, F.\inst{3} \and Lovis, C.\inst{3} \and H\'ebrard, G.\inst{1}  }
\institute{Institut d'Astrophysique de Paris, Universit\'e Pierre et Marie Curie, UMR7095 CNRS, 98bis bd. Arago, 75014 Paris, France \and Observatoire de Haute Provence, CNRS/OAMP, 04870 St Michel l'Observatoire, France \and Observatoire de Gen\`eve, Universit\'e de Gen\`eve, 51 Ch. des Maillettes, 1290 Sauverny, Switzerland}
\abstract{
 For fiber-fed spectrographs with a stable external wavelength source, scrambling properties of optical fibers and, homogeneity and stability of the instrument illumination are important for the accuracy of radial-velocimetry. Optical cylindric fibers are known to have good azimuthal scrambling. In contrast, the radial one is not perfect. In order to improve the scrambling ability of the fiber and to stabilize the illumination, optical double scrambler are usually coupled to the fibers. Despite that, our experience on SOPHIE and HARPS has lead to identified remaining radial-velocity limitations due to the non-uniform illumination of the spectrograph. We conducted tests on SOPHIE with telescope vignetting, seeing variation and centering errors on the fiber entrance. We simulated  the light path through the instrument in order to explain the radial velocity variation obtained with our tests. We then identified the illumination stability and uniformity has a critical point for the extremely high-precision radial velocity instruments (ESPRESSO@VLT, CODEX@E-ELT). Tests on square and octagonal section fibers are now under  development and SOPHIE will be used as a bench test to validate these new feed optics. 
} 
\maketitle
\section{Introduction}
\label{intro}
High-precision radial-velocimetry (RV) is still a main technique in the search and characterization of planetary systems. Up to date, most of the known extrasolar planets have been detected with this method. It will remain at the forefront of exoplanet science for the coming years thanks to its efficiency in finding low-mass planets since HARPS has reached an RV accuracy less than 1ms$^{-1}$ (ref.~\cite{RefM}). In the aim to reach Earth-type planets and, also measure the expansion of the universe, new spectrographs in the visible are now in study to reach a precision of only few cms$^{-1}$ in the near future (ESPRESSO@VLT \cite{RefPa1}, CODEX@E-ELT \cite{RefPa2}). 

To reach this level of accuracy, we must avoid a certain number of limitations. Among these, one can quote the stellar noise (activity, pulsation, surface granulation), the contamination by external sources (moon, close-by objects, background continuum), and what light may encounter during its path (e.g. atmospheric dispersion). As part of instrumental factors, CCD  cosmetics, charge transfer inefficiency (ref.~\cite{RefB1}), wavelength calibration (refs.~\cite{RefL} and~\cite{RefC}) should be mentioned. For an exhaustive discussion of these limitations, we refer to~\cite{RefP}. The authors highlight the impact of the spectrograph illumination as one of the main limiting factors, and the need to analyze and quantify it. In this proceeding, we then focus on the limitation due to the injection of light in the fiber-fed spectrograph. This work is based on our experiment on two high-resolution, high-precision fiber-fed spectrographs, SOPHIE@1.93m, Observatoire de Haute-Provence (OHP), France (refs.~\cite{RefB2} and ~\cite{RefS}) and HARPS@3.6m, European Southern Observatory (ESO)-La Silla, Chile (ref.~\cite{RefM}). 

Up to day, SOPHIE has a Doppler accuracy of about 5~ms$^{-1}$, as most of high-accuracy spectrographs involved in exoplanet survey. In service since November 2006, we identified that instrumental limitations 
mainly come from the old Cassegrain Fiber Adaptater (developed for ELODIE spectrograph). In order to reach the so far unique HARPS level of accuracy, we conducted tests that illustrate the incomplete fiber scrambling. Detected with SOPHIE, these effects were also characterized on HARPS at a lower level. In the development of the future instruments, the optimization of SOPHIE is a bench test for the improvement of the spectrograph illumination uniformity and stability. 

The proceeding is structured as followed. In a the second section, we expose the properties of the fiber-fed spectrographs and the light injection. The third section reviews the different effects that are detected in the RV due to the incomplete and imperfect scrambling obtained with the standard fibers. In the last section, we describe solutions proposed to optimize the fibers scrambling.

\section{Fiber-fed spectrographs}
\label{sec:1}

HARPS and SOPHIE are fiber-fed spectrographs with simultaneous reference calibrations. The stellar light collected by the telescope are lead to the instrument through a standard step-index multi-mode cylindrical optical fiber. SOPHIE has two observing mode using two different fibers, the High-Resolution (HR) and the High-Efficiency mode (HE). In HR mode, the spectrograph is fed by a 40.5-$\mu$m slit superimposed on the 
output of the 100-$\mu$m fiber, reaching a spectral resolution of $\lambda/\Delta\lambda$ = 75,000. In HE mode, the spectrograph is directly fed by the 100-$\mu$m fiber with a resolution power of 40,000. Both SOPHIE fibers have an sky acceptance of 3-arcsec. HARPS has 70-$\mu$m fiber with a sky acceptance of 1-arcsec and a spectral resolution of 110,000 (see Table~\ref{tab:1}).
 
Non-uniform illumination of the slit or output fiber at the spectrograph entrance decreases the radial-velocity precision. Indeed, variations in seeing, focus and image shape at the fiber entrance may induced non-uniform illumination inside the pupil of the spectrograph. The optical aberrations lead to variations in the centroids of the stellar lines on the focal plane. \\

\begin{table}[b]
\caption{Comparison of parameters of fiber-fed spectrographs {\bf SOPHIE and HARPS}.}
\label{tab:1}      
\begin{tabular}{cccc}
\hline\noalign{\smallskip}
Instrument & Doppler prec.  & 1 pixel & Resol. $\Delta\lambda/\lambda$ \\
HARPS & $\leqslant$ 1ms$^{-1}$ & 800 ms$^{-1}$ & 110,000 \\
SOPHIE (HR) & $\sim$ 4 ms$^{-1}$& 1400 ms$^{-1}$ & 75,000 \\
SOPHIE (HE) & $\sim$ 10 ms$^{-1}$  &   1400 ms$^{-1}$ & 40,000  \\
\noalign{\smallskip}\hline
\end{tabular}
\end{table}

\begin{figure}
\resizebox{0.9\columnwidth}{!}{
  \includegraphics{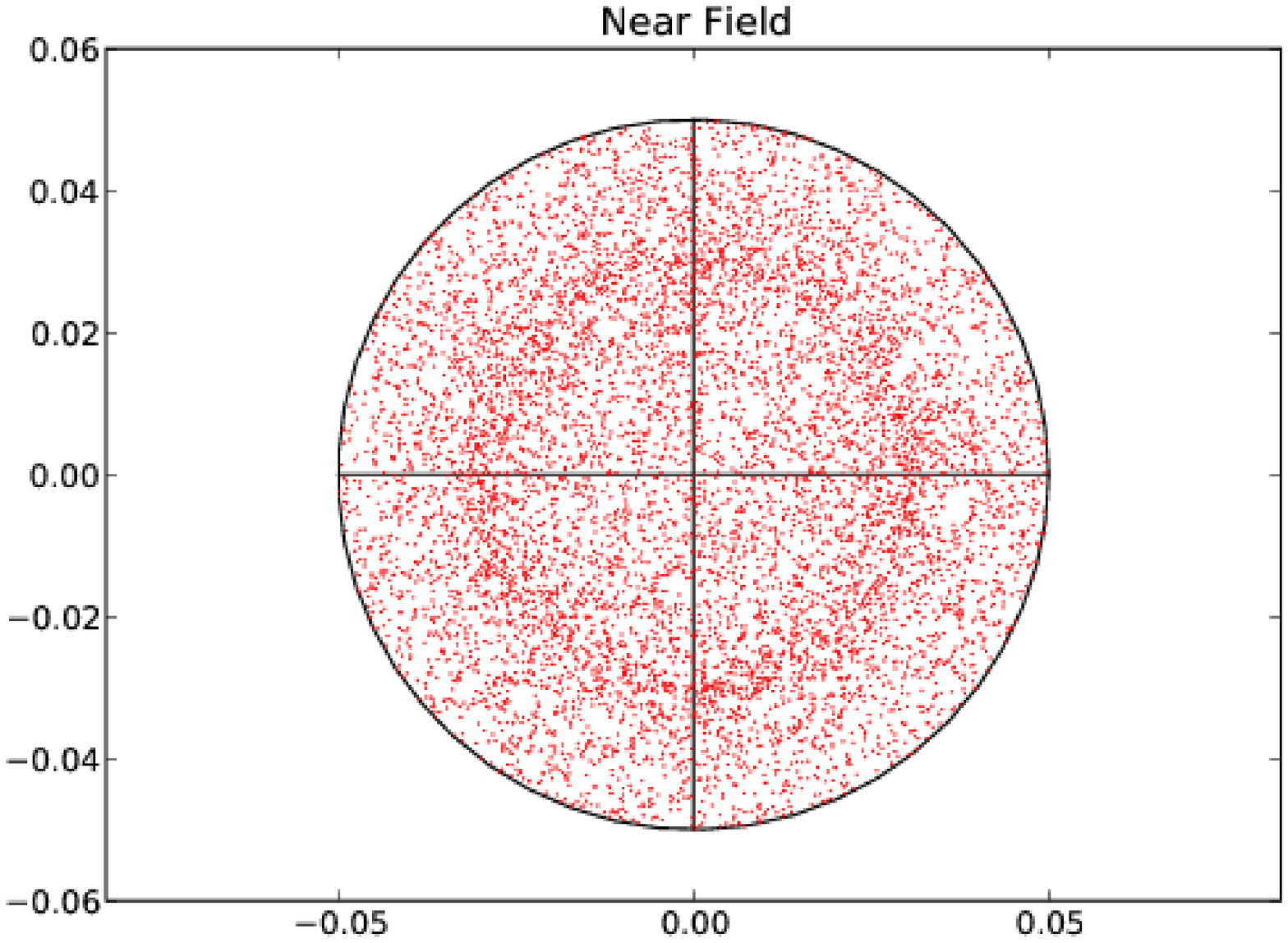} }
   \resizebox{0.9\columnwidth}{!}{ \includegraphics{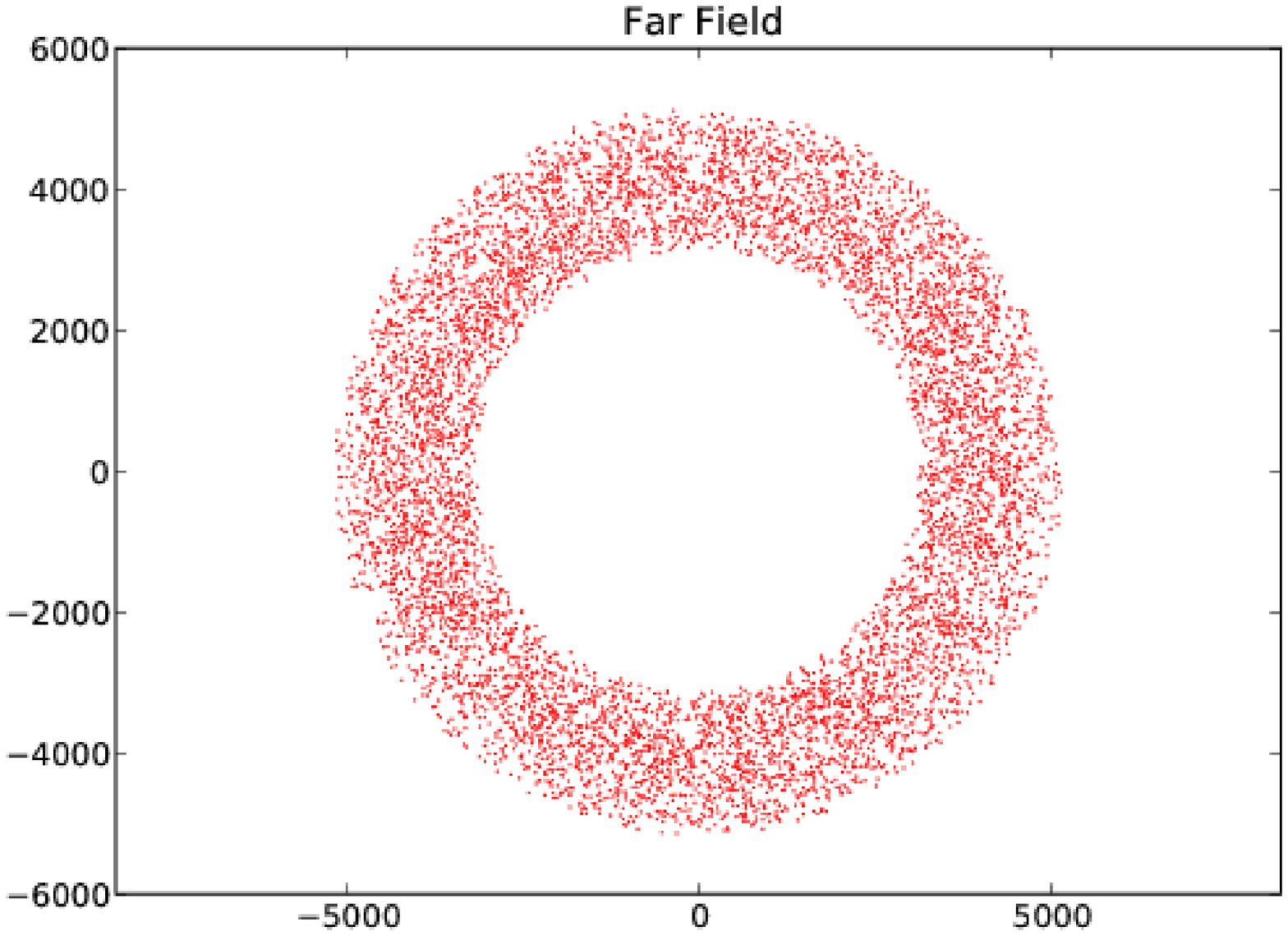} }
\caption{Illumination at the output of one cylindrical optical fiber in the near and the far field. Sources are off axes. A high degree of azimuthal scrambling is observed. In contrast, it remains some effects in the radial one. }
\label{fig:3}  
\end{figure}

Optical fibers are used to lead the light from the telescope to the entrance of the spectrograph. They have the properties to scrambler the atmospheric effects and guiding and centering errors discussed previously. But the scrambling ability of one multimode fiber is not perfect as shown in Fig.~\ref{fig:3}. With an input off axes image, the azimuthal scrambling is good whereas it remains some effects in the radial one.
This pattern observed is bigger in far field than in near field. Near field and far field are respectively defined as, the brightness distribution across the output face of the fiber, and, as the angular distribution of light of fiber output beam. Therefore, changes in the input beam cause only circular symmetric changes in the fiber output pattern. In contrast, imperfect radial scrambling allows small zonal errors to remain. \\

 HARPS and the SOPHIE HR mode are equipped with optical double scramblers. It is used to increase and improve the uniformity and stabilization of the illumination of the spectrograph entrance (refs.~\cite{RefBr} and~\cite{RefH}). The near field of a circular fiber is observed to be better scramble than the far field. In spite of only one fiber guiding the light from the telescope to the instrument, double scrambler is composed of two fibers coupled with two doublets. The system is designed to inverse near field and far field in order to induce a better radial scrambling, although it causes flux losses (ref.~\cite{RefB0}).

\section{RV {\bf effects} of fiber {\bf imperfect} scrambling}

\subsection{Centering/Guiding on the fiber entrance}

The degree of radial scrambling describes the stability of the output beam as the input image is moved from the center to the edge of the fiber. It is possible to detect in RV some variations due to the insufficiency of radial scrambling. The light from a stable star is moved with the guiding/centering system from an edge to the other of the fiber. We compute the RV with the data reduction pipeline for different positions of the star. The test was done for the two modes of SOPHIE and for HARPS. The results are computed in Table~\ref{tab:2}.

\begin{table}[b]
\caption{Variation observed in RV as the input image (star) is moved from the center to the edge of the fiber. HARPS and SOPHIE with High-Resolution fiber used a double scrambler instead of the High-Efficiency mode of SOPHIE.}
\label{tab:2}      
\begin{tabular}{ccc}
\hline\noalign{\smallskip}
Instrument & fib. diam. & RV effect [ms$^{-1}$]\\
HARPS & 1" & $\sim$ 3  \\
SOPHIE (HR) & 3" & $\sim$ 13 \\
SOPHIE (HE) & 3" & $\sim$ 36 \\
\noalign{\smallskip}\hline
\end{tabular}
\end{table}

 Because the scrambling of the fiber is not perfect, the movement of the input image corresponds to a shift on the detector. We observed first that, wider is the fiber, more important is the RV variations.
Moreover, the value of meter per second per pixel vary as a function of the resolution power. Moving the input image from the center to the edge is then expected to have a lower effect for a higher spectral resolution. In addition, we expect that the double scrambler improve the radial scrambling, i.e. decrease the RV effect. 

\subsection{Vignetting telescope}

\begin{figure}
\resizebox{0.9\columnwidth}{!}{
  \includegraphics{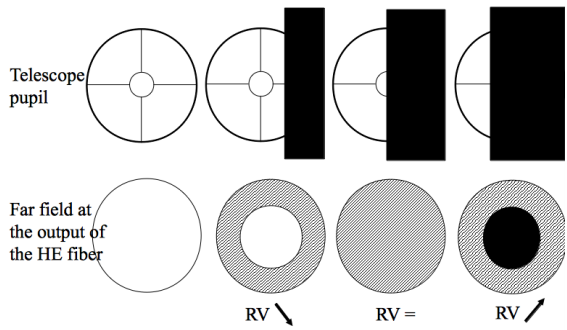} }
\caption{Schema illustrating how the vignetting of the telescope is translated at the output of the fiber HE in the far-field, i.e. at the entrance of the spectrograph. For external occultation, the RV decrease whereas, for internal occultation the RV increase.}
\label{fig:4}  
\end{figure}

The pupil of the telescope, i.e. the entrance of the telescope or the far field of the input light is known to be more stable than the telescope image. We would like to test the effect in RV of variations of the far field. For that, we vignette the telescope on day sky with the dome. We computed the RV with the pipeline SOPHIE, estimating that the variation of Barycentric Earth RV is not significant during our measurements. We made this test with the HE mode in order that the variation of the far-field of the telescope are projected on the grating (i.e. without double scrambler). We remarked that the way of the RV variations depends on exterior or interior vignetting (Fig.~\ref{fig:4}).

\subsection{Optical path simulations}

\begin{figure}
\resizebox{1.\columnwidth}{!}{
  \includegraphics{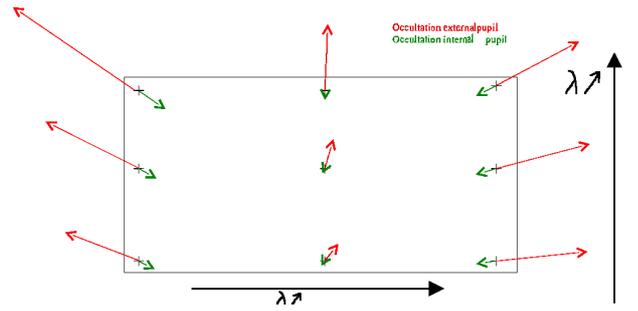} }
\caption{Displacement of the position of a nine wavelength as a function of a internal (green) or an external (red) occultation of the entrance of the spectrograph on the CCD detector.}
\label{fig:2}  
\end{figure}

We simulated the optical path in the SOPHIE spectrograph. We observed a displacement of the spectrum in function of  wavelength on the focal plane when the input beam at the entrance of the spectrograph is center-illuminated or center-darked, corresponding respectively when vignetting external and internal part of the pupil in HE mode (cf. Sect.3.2). Variation of the slit or the optical fiber illumination are directly translated on the spectrum in Fig.~\ref{fig:2}. We observed that the shift is more significant for an external occultation. Because the effect is not symmetric along an order and not monitored by the calibration lamp, the final computed RV vary. So, variations of the far-field pattern projected onto the grating can introduce radial-velocity shifts at the detector lowering the final precision in RV.

\begin{figure}
\resizebox{1.\columnwidth}{!}{
  \includegraphics{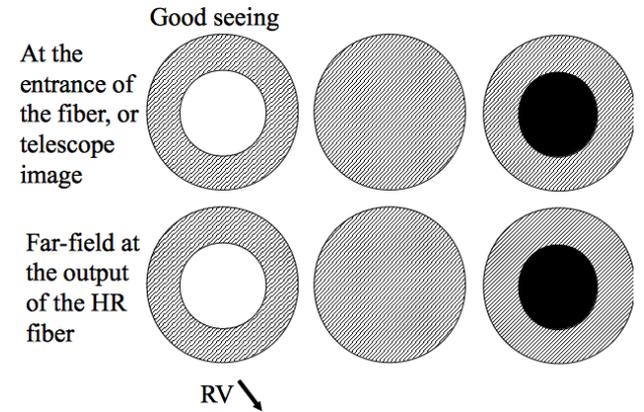} }
\caption{When the seeing is good, the telescope image is less wide. With the double scrambler in HR mode, near field becomes far-field and is projected on the collimator at the entrance of the spectrograph. The pattern is comparable to that observed in HE when vignetting telescope (Fig.~\ref{fig:4})}
\label{fig:6}  
\end{figure}

\subsection{RV effects due to seeing variations}
\begin{figure}[t]
\centering
\resizebox{0.95\columnwidth}{!}{
  \includegraphics{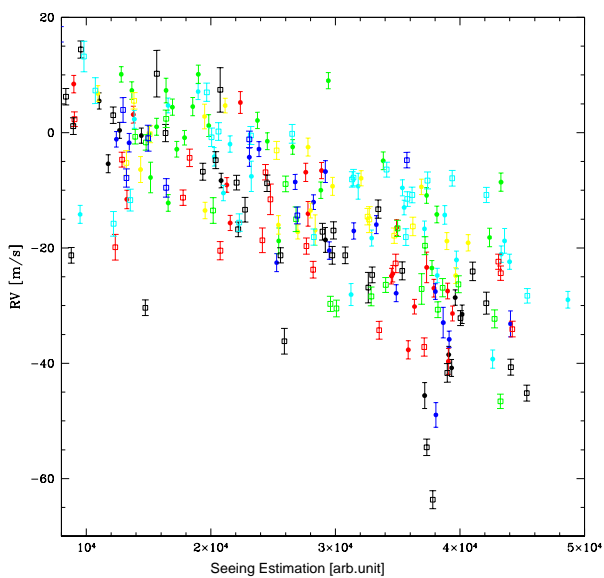} }
\caption{Radial-velocity measurements as a function of a seeing index (cf. text). The seeing decrease when the seeing estimation $S$ increase. As the seeing is lower, the RV decreases.}
\label{fig:1}     
\end{figure}

Variations of the far-field of the input beam in HE mode is equivalent to a variation of the telescope image in HR mode, due to the double scrambler. Variations in seeing induce a variation of the input image as illustrate in Fig.~\ref{fig:6}.
The input beam at the entrance of the spectrograph is center-illuminated (external occultation) when a target is observed with a good seeing in HR mode. The previous  tests and simulations explain a systematic effect observed with SOPHIE named as "seeing effect". For a sample of stars observed at high signal to noise ratio (SNR) in HR mode, we observed a decrease in RV correlated with an estimation of seeing. The value of the seeing is not monitored with SOPHIE. We estimated it with calculating the relative flux by unit of exposure time :
\begin{equation}
S = \frac{SNR^{2}}{T_{exp} 10^{-M_{V}/2.5}}
\end{equation} 
with SNR, the signal to noise ratio of the spectra, T$_{exp}$ the time exposure of the measurement, M$_{V}$ the visual magnitude of the target. The seeing decrease when the seeing estimation $S$ increase.
 With a lower seeing, the input image is smaller than the fiber width, and as shown in the simulations, induced a variation of the far field input in the spectrograph, due to the double scrambler. We expect and observed a decrease of RV when the value of the seeing decrease as shown in Fig.~\ref{fig:1}.

Moreover, when the input image is smaller than the diameter of the fiber, we are more sensible to effect of guiding and centering system. The guiding and centering system may induced variation of the position of the input image on the slit or the entrance of the fiber, and so introduce RV variations. These RV variation are quite random, and add some noise.

Whereas the seeing effect is directly related to the optical path in the spectrograph, it may be quantified. We are modeling this variations with our data in order to remove this noise with a software tool. 

We note that the effect on HARPS is expected to be lower. The 1" width of the fibers is smaller than the SOPHIE and roughly equivalent to the typical seeing at La~Silla. Furthemore the quality image provided by the 3.6-m telescope is close to 0.7 arcsec. Hence, HARPS is less sensible to the variations of seeing. Moreover, the resolution power of the instrument is better and then the incomplete radial scrambling has a lower impact in RV as shown in Sect.~3.1. Furthermore the optical aberrations on HARPS are smaller than in SOPHIE.

\section{Perspectives and conclusions}

The CODEX experiment calls for a Doppler precision as low as 1 cms$^{-1}$. This work shows that the future instruments need for better scrambling to reach this accuracy. New types of fibers for feeding the telescope or new double scrambler are needed to improve the precision and to minimize the noise due to variations of the illumination at the entrance of the spectrograph. 
To remove the spherical symmetry of cylindrical fibers, octagonal and square section optical fibers are then in study at Geneva, OHP-LAM, and ESO Garching. As first results, very good scrambling properties are observed in the near field, whereas strange patterns in far field are not well understood at the moment. It needs future tests to improve the measurements, to study its best used, on all the path or in a double scrambling mode. SOPHIE is then going to be a bench test for the future instruments.

We note that additional solution is being installed on HARPS to reduce the guiding and centering limitations. A tip-tilt mirror generate small-amplitude and high frequencies variations in order to induce a spread of the light. 

 A new guiding camera is now in operation on SOPHIE and allows a better guiding and centering and 
also allows to monitor the exact seeing.

\end{document}